\documentclass[aps,pre,preprint,superscriptaddress,showpacs]{revtex4}
\usepackage{bm}
\begin{document}

\title{Berry Effect in Unmagnetized Inhomogeneous Cold Plasmas}

\author{Reza Torabi}
\email{rezatorabi@aut.ac.ir} \affiliation{Physics Department,
Tafresh University, P.O.Box: 39518-79611, Tafresh, Iran}
\author{Mohammad Mehrafarin}
\email{mehrafar@aut.ac.ir} \affiliation{Physics Department,
Amirkabir University of Technology, Tehran 15914, Iran}

\begin{abstract}
The propagation of electromagnetic waves in an unmagnetized weakly
inhomogeneous cold plasma is examined. We show that the
inhomogeneity induces a gauge connection term in wave equation,
which gives rise to Berry effects in the dynamics of polarized rays
in the post geometric optics approximation. The polarization plane
of a plane polarized ray rotates as a result of the geometric Berry
phase, which is the Rytov rotation. Also, the Berry curvature causes
the optical Hall effect, according to which, rays of left/right
circular polarization deflect oppositely to produce a spin current
directed across the direction of propagation.
\end{abstract}


\maketitle

\section{Introduction}

It has often been said that ninety nine percent of the matter in the
universe is in plasma state \cite{Chen}. Because real plasmas have
density inhomogeneity, the study of electromagnetic wave propagation
in inhomogeneous plasmas has become important
\cite{Hacquin,Cho,Malik}. In this paper we examine the propagation
of electromagnetic waves in an unmagnetized plasma via the post
geometric optics approximation \cite{Bliokh,Bliokh1}. The plasma is
assumed to have weak stationary density inhomogeneity, and thermal
motions are not considered. Different polarizations, the left and
right circular polarizations, are degenerate in a homogeneous
isotropic plasma medium \cite{Krall}. In the presence of
inhomogeneity, we show that this double degeneracy is lifted by a
gauge connection term in the wave equation, which gives rise to
Berry effects in the dynamics of polarized rays in the post
geometric optics approximation. The polarization plane of a plane
polarized wave rotates as a result of the geometric Berry phase,
which is the Rytov-Vladimirskii rotation \cite{Rytov,Vladimirskii}.
This is, of course, in contrast to the well known Faraday rotation,
which is a dynamical effect due to the interaction of light and the
magnetic field in a medium. Berry phase is a non-integrable phase
factor arising from the adiabatic transport of a system around a
closed path in its parameter space \cite{Berry}. Geometrically, it
originates from parallel transport in the presence of a gauge
connection in the parameter space \cite{Simon}. When an
electromagnetic wave travels in a weakly inhomogeneous medium, the
direction of the wave vector varies slowly so that the parameter
space in this case corresponds to the momentum space.  The Berry
curvature associated with the gauge connection in momentum space
causes the optical Hall (or Magnus) effect \cite{Dooghin,Liberman},
according to which, waves of left/right circular polarization
deflect oppositely to produce a spin (polarization) current directed
across the direction of propagation. Such Berry effects are very
typical of spin transport and have been derived repeatedly for
various particles, in particular, for photons in different
inhomogeneous media
\cite{Bliokh,Bliokh1,Bliokh2,Onoda,Onoda2,Sawada,Torabi,Torabi2}.

The paper is organized as follows. In section II, we write the
plasma wave equation in an operator form by introducing a
`Hamiltonian' operator and show that, in the process of its
diagonalization, a gauge connection emerges in the momentum space
because of the inhomogeneity. In section III, we present the
post-geometric approximation which is suitable for weak
inhomogeneity, and focus on the circularly polarized states by
projecting the Hamiltonian on the polarization subspace. Finally, in
section IV, we derive the Berry effects in the dynamics of polarized
rays by considering the post-geometric optics Hamiltonian, and
establish the Rytov and the optical Hall effects in the plasma
medium.

\section{Plasma wave equation and the gauge connection}

Consider an inhomogeneous plasma which can be generated by gravity
or a position dependent electric field. Generally, both of them can
exist and the balance equation of the forces in steady state is
\[
\rho_{\alpha 0}({\bm x})q_\alpha {\bm E}_0 ({\bm
x})+\rho_{\alpha0}({\bm x}){\bm F}_{g\alpha}-m_\alpha \nabla
p_{\alpha0}=0,
\]
where $\rho_{\alpha 0}$ is unperturbed mass density, ${\bm F}_g$ is
the gravitational force and $\alpha=e,i$ represent electrons and
ions. Now, we want to derive the equation which determines the wave
properties of an inhomogeneous unmagnetized plasma. The procedure is
similar to the derivation of the wave equation in homogeneous plasma
\cite{Krall} but we should notice that the density $\rho_{\alpha
0}(\bm{x})$ is now a function of $\bm{x}$. Considering small
harmonic perturbation about steady state
\[
\rho_{\alpha}(\bm{x})=\rho_{\alpha0}(\bm{x})+\rho_{\alpha1}(\bm{x})e^{-i\omega
t},
\]
\[
\bm{V}_{\alpha}=\bm{V}_{\alpha1}(\bm{x})e^{-i\omega t},
\]
\[
\bm{E}={\bm E}_0({\bm x})+\bm{E}_{1}(\bm{x})e^{-i\omega t},
\]
\[
\bm{B}=\bm{B}_{1}(\bm{x})e^{-i\omega t},
\]
we linearize the two fluid and Maxwell's equations. The linearized
momentum equation is
\[
-i\omega \rho_{\alpha 0}({\bm x}){\bm V}_{\alpha1}({\bm
x})=\frac{q_\alpha \rho_{\alpha0}({\bm x})}{m_\alpha}{\bm E}_1({\bm
x})+\rho_{\alpha1}({\bm x})\frac{\nabla p_{\alpha0}}{
n_{\alpha0}({\bm x})}-\gamma\nabla \bigg{(}\frac{\rho_{\alpha1}({\bm
x})p_{\alpha0}}{\rho_{\alpha0}({\bm x})} \bigg{)}.
\]
In deriving the above, the balance equation and $p_{\alpha1}=\gamma
p_{\alpha0}\rho_{\alpha 1}({\bm x})e^{-i\omega t}/\rho_{\alpha
0}({\bm x})$ have been used. After combining it with linearized
Maxwell's equations, the wave equation will be
\[
\nabla\times\nabla\times
\bm{E}_{1}-k_{0}^{2}\bigg{(}1-\frac{\omega_{pe}^{2}({\bm
x})}{\omega^{2}}-\frac{\omega_{pi}^{2}({\bm
x})}{\omega^{2}}\bigg{)}\bm{E}_{1}=\frac{4\pi}{c^2}\sum_\alpha
\frac{q_\alpha}{m_\alpha}\bigg{[}\gamma\nabla
\bigg{(}\frac{\rho_{\alpha 1}({\bm x})p_{\alpha0}}{\rho_{\alpha
0}({\bm x})}\bigg{)}-\frac{\rho_{\alpha 1}({\bm x})}{\rho_{\alpha
0}({\bm x})}\nabla p_{\alpha0}\bigg{]},
\]
where $k_{0}=\omega/c$ and $\omega_{p\alpha}^{2}({\bm x})=4\pi
\rho_{\alpha 0}({\bm x})q_{\alpha}^2$. The cold plasma approximation
and assuming stationary and inhomogeneous density is usually used to
describe wave propagation in inhomogeneous media
\cite{Ginzburg,Hacquin}. Under the cold plasma approximation, we can
drop the right hand side of the equation and it takes the form
\begin{equation}
\label{1} \nabla\times\nabla\times
\bm{E}_{1}=k_{0}^{2}\bigg{(}1-\frac{\omega_{pe}^{2}({\bm
x})}{\omega^{2}}-\frac{\omega_{pi}^{2}({\bm
x})}{\omega^{2}}\bigg{)}\bm{E}_{1}.
\end{equation}

Introducing
\[
\epsilon(\bm{x},\omega)=\bigg{(}1-\frac{\omega_{pe}^{2}({\bm
x})}{\omega^{2}}-\frac{\omega_{pi}^{2}({\bm
x})}{\omega^{2}}\bigg{)}\equiv n^{2}(\bm{x},\omega),
\]
and the dimensionless 'momentum' operator \cite{Bliokh1,Bliokh2},
$\bm{p}=-ik_{0}^{-1}\nabla$, (1) reads
\begin{equation}
\label{2} H\bm{E}_{1}=0,
\end{equation}
where the 'Hamiltonian' $H$ is a matrix-valued differential operator
with elements
\[
H_{ij}(\bm{x},\bm{p},\omega)=[{\bm{p}}^{2}-n^{2}(\bm{x},\omega)]\delta_{ij}-p_{i}p_{j}.
\]
In inhomogeneous plasma media the refractive index is also a
function of frequency as well as coordinate. The refractive index is
real for frequencies approximately above the plasma frequency and is
imaginary for frequencies below it which causes the first to remain
and the later to be absorbed in the media.

The momentum operator obeys the standard commutation relations
\[
[x_{i},p_{j}]=ik_{0}^{-1}\delta_{ij}.
\]
which has an important consequence with regard to diagonalization of
$H$. To this end, we build a unitary matrix $R(\bm{p})$ from the
eigenvectors of the non-diagonal part $p_{i}p_{j}$, according to
\begin{eqnarray}
R(\bm{p})=\left( \begin{array}{ccc} \frac{p_y }{\sqrt {p_x^2 +p_y^2
}} & -\frac{p_x p_z }{p\sqrt{p_x^2 +p_y^2 }} & \frac{p_x }{p}
\\ -\frac{p_x }{\sqrt {p_x^2 +p_y^2 }} &
-\frac{p_y p_z }{p\sqrt{p_x^2 +p_y^2 }} & \frac{p_y }{p} \\
0 & \frac{\sqrt {p_x^2 +p_y^2 } }{p} & \frac{p_z }{p} \end{array}
\right)
\end{eqnarray}
where $p=|\bm{p}|$, of course. The unitary transformation
$H\rightarrow R^{\dag}HR$, $\bm{E}_{1}\rightarrow R^{\dag}
\bm{E}_{1}$ applied to equation (2), then yields
\begin{equation}
\label{4}
H (\bm{x},\bm{p},\omega)=\bm{p}^2 1-\Lambda -{R}^{\dag}n^2
R,
\end{equation}
where $\Lambda=diag(0,0,\bm{p}^{2})$ and 1 stands for the unit
matrix. Of course, if $\bm{x}$ and $\bm{p}$ were classical
(commuting) variables, the 'potential' part of the Hamiltonian (5)
would simply reduce to $n^{2} 1$, and $H$ would be completely
diagonalized. However, in view of their anti-commutation, we have
for the potential term in the momentum representation,
\[
R^{-1}({\bm p})n^{2}(ik_{0}^{-1}\nabla_{{\bm p}}) {R}({\bm
p})=n^2(1ik_{0}^{-1}\nabla_{{\bm p}}-k_{0}^{-1}{\bm A})=n^2({\bm
x}1-k_{0}^{-1}{\bm A}),
\]
where
\[
i{\bm A}({\bm p})=R^{-1}\nabla_{{\bm p}}R.
\]
In deriving the above, we have made use of the identity
\cite{Bliokh2,Torabi2}
\[
G^{-1}(x)f(\partial_x)G(x)=f(\partial_x+G^{-1}\partial_x G),
\]
since $R^{-1}R$=1, the vector matrix ${\bm A}$ is Hermitain. Thus
\[
H({\bm x},{\bm p},\omega)={\bm p}^2 1-\Lambda -n^2({\bm
x}1-k_{0}^{-1}{\bm A}).
\]
The canonical (generalized) coordinate conjugate to ${\bm p}$ is
${\bm x}$ which corresponds to the usual derivative
$ik_0^{-1}\nabla_{{\bm p}}$. In the absence of inhomogeneity, this
canonical coordinate is the physical (observable) coordinate. With
inhomogeneity present, however, the new coordinate, ${\bm
x}1-k_0^{-1}{\bm A}$, corresponds to $ik_0^{-1}{\cal D}_{{\bm p}}$,
where ${\cal D}_{{\bm p}}$ is the covariant derivative defined by
\[
{\cal D}_{{\bm p}}=1\nabla_{{\bm p}}+i{\bm A}({\bm p}).
\]
The inhomogeneity can be, thus, viewed as inducing the non-Abelian
gauge connection (potential) ${\bm A}({\bm p})$ in the momentum
space. Such a gauge potential is a well known feature of spin
transport \cite{Bliokh1,Bliokh2,Torabi2,Mehrafarin,Berard}. $\bm{A}$
is a \textit{pure} gauge potential, i.e., the corresponding field
strength (curvature), $\nabla_{{\bm p}}\times{\bm A}+i{\bm A}\times
{\bm A}$, is identically zero. The covariant derivatives, thus,
commute so that the new coordinates also commute. Furthermore, the
latter satisfy the same commutation relations with ${\bm p}$ as the
canonical coordinates ${\bm x}$.

We consider that the inhomogeneity is weak, thus the direction of
the wave propagation varies slowly (infinitely slowly for adiabatic
density variations). Let us take this direction to be along the
$z$-axis so that $E_{1z}$ is small (negligible in the adiabatic
limit) compared to the other two components. Since we are
considering polarization transport, we introduce the unitary matrix
\[
V=\frac{1}{\sqrt 2 } \left({{\begin{array}{ccc}
 1 & 1 & 0 \\
 i & -i & 0\\
 0 & 0 & \sqrt{2}\\
\end{array} }} \right)
\]
and write the wave equation in the helicity basis via the unitary
transformation
\begin{equation}
H\rightarrow V^\dag H V={\bm p}^2 1-\Lambda -n^2({\bm
x}1-k_0^{-1}\tilde{{\bm A}}),\ \ {\bm E_1}\rightarrow V^\dag{\bm
E_1}=(E_{1+},E_{1-},E_{1z}) \label{ham}
\end{equation}
where $\tilde{\bm{A}}=V^\dag \bm{A} V$, and
$E_{1\sigma}=(E_{1x}-i\sigma E_{1y})/\sqrt{2}$ represents the two
circularly polarized states with helicity $\sigma=\pm 1$. The
elements $i,j=1,2$ of $H$, thus, correspond to the polarization
subspace, on which we shall focus.

\section{The post-geometric approximation}

We now proceed to study the dynamics imposed by the Hamiltonian
(\ref{ham}) based on the Taylor expansion of $n^2({\bm
x}1-k_0^{-1}\tilde{{\bm A}})$. $k_0^{-1}$ serves as a small
parameter for the expansion, provided it is smaller than the length
scale, $L$, of the density variations in the plasma. The zeroth
order approximation (the geometric approximation), $k_0^{-1}\ll L$,
thus holds appropriate for large $L$, i.e., for a practically
homogeneous plasma. In the zeroth approximation the Hamiltonian
(\ref{ham}) becomes diagonal,
\[
H^{(0)}({\bm x},{\bm p},\omega)=diag({\bm p}^{2}-n^{2},{\bm
p}^{2}-n^{2},-n^{2}),
\]
which implies a double degeneracy in the polarization subspace:
electromagnetic waves with left/right circular polarizations have
the same dispersion in a homogeneous isotropic plasma \cite{Krall}.
Furthermore, $E_{1z}=0$, and the wave travels unrefracted along the
$z$-direction, as expected.

In inhomogeneous plasmas, the refractive index gradient or
equivalently the plasma frequency gradient removes the polarization
degeneracy. In the first approximation (the post-geometric
approximation), which is suitable for weak inhomogeneity, the
Hamiltonian (\ref{ham}) takes the form
\[
H^{(1)}({\bm x},{\bm p},\omega)=H^{(0)}({\bm x},{\bm
p})+k_0^{-1}\nabla n^2\cdot\tilde{{\bm A}}.
\]
Calculating $\tilde{\bm{A}}$ via Appendix A, the non-diagonal
correction term is seen to couple $E_\sigma$ to $E_{1z}$, too. Since
$E_{1z}$ is small (negligible in the adiabatic limit) this coupling
can be ignored. Consequently, we can project $H^{(1)}$ on the
polarization subspace with the result
\[
H^{(1)}({\bm x},{\bm p},\omega)=({\bm p}^2-n^2)1+k_0^{-1}\nabla
n^{2}.{\bm A}_{\bot}\sigma_3,
\]
where ${\bm A}_{\bot}({\bm
p})=\frac{p_3}{p(p_1^2+p_2^2)}(-p_2,p_1,0)$ and $\sigma_3$ is the
Pauli matrix. The wave equation, thus, breaks down into two
independent equations for the left/right circularly polarized waves
according to $\mathcal{H}_\sigma E_\sigma=0$, where
\begin{equation}
\label{eq}
\mathcal{H}_\sigma=\frac{1}{2}[({\bm p}^{2}-n^{2}({\bm
r},\omega)]=\frac{1}{2}({\bm p}^{2}-n^{2}+\sigma k_0^{-1}\nabla
n^{2}.{\bm A}_{\bot}).
\end{equation}
(The factor $\frac{1}{2}$ has been introduced for later
convenience.) As remarked, the refractive-index gradient is
responsible for the removal of polarization degeneracy. ${\bm
A}_{\bot}\sigma_3$, or equivalently its eigenvalue
$\bm{A}_{\bot}\sigma$, is the (Abelian) gauge potential that emerges
in the momentum space in the adiabatic approximation. The
corresponding field strength (Berry curvature) and the physical
coordinates are, thus, given by $\nabla_{\bm p}\times {\bm A}_{\bot}
\sigma=-p^{-3} {\bm p} \sigma$ and ${\bm r}={\bm x}-k_0^{-1}{\bm
A}_{\bot} \sigma$ respectively. Note that because of the
non-vanishing gauge field strength,  which is the field of a
magnetic monopole of charge $-\sigma$ situated at the origin of
momentum space, the physical coordinates, now, do not commute:
\[
[r_i,r_j]=i\sigma k_0^{-2}\varepsilon_{ijk}\frac{p_k}{p^3}.
\]

\section{Berry effects in the dynamics of polarized rays}

Having constructed the Hamiltonian, the equations of motion of an
electromagnetic ray can be obtained via the Hamilton's equations
\cite{Kravtsov}. In the post geometric approximation, the
semi-classical equations of motion of a circularly polarized ray
are, therefore,
\[
\dot{\bm p}=-\nabla_{\bm x}\mathcal{H}_\sigma,\;\;\; \dot{\bm
x}=\nabla_{\bm p}\mathcal{H}_\sigma,
\]
where dot denotes derivative with respect to the ray parameter $s$,
defined in terms of the ray length, $l$, by $dl=n ds$. The physical
coordinate ${\bm r}={\bm x}-k_0^{-1}{\bm A}_{\bot} \sigma$ and the
momentum ${\bm p}=k_0^{-1}{\bm k}$ are now considered classical, of
course. Along the trajectory, they represent the ray's position and
(dimensionless) wave vector, respectively. Thus, using (\ref{eq}),
\begin{equation}
\label{10}
\dot{\bm p}=\frac{1}{2}\nabla_{\bm r}n^2,\;\;\;{\bm
\dot{\bm r}}={\bm p}+\sigma k_0^{-1}\frac{\bm {p}\times \dot{\bm{p}}
}{p^3}.
\end{equation}
These, of course, reduce to the standard ray equations of geometric
optics in the zeroth (`classical') approximation,
$k_0^{-1}\rightarrow 0$, where the left/right circularly polarized
rays follow the same trajectory. However, in the post-geometric
(semi-classical) approximation, as seen from (\ref{10}), the rays
split due to the effect of Berry curvature of the momentum space
(the magnetic monopole-like gauge field strength). The deflections
from their classical (geometric optic) trajectories are given by
\[
\delta{{\bm r}}= \sigma k_0^{-1}\int_C \frac{{\bm p}\times d{\bm
p}}{p^3},
\]
where $C$ is the ray trajectory in momentum space. The resulting
displacements are, therefore, opposite and locally orthogonal to the
direction of propagation. This, which is a general feature of spin
transport, constitutes the optical Hall effect in inhomogeneous
plasmas.

The phase change suffered by the ray in the course of its
propagation is given by
\[
\phi=\omega t-\int {\bm k}\cdot d{\bm x}=\omega t-k_0\int {\bm
p}\cdot d{\bm r}+ \sigma \int_C {\bm A}_{\bot} \cdot d{\bm p},
\]
where in the last integral, we have used the fact that ${\bm
A}_{\bot}\cdot {\bm p}=0$. This integral represents the geometric
Berry phase, which is of opposite signs for the two polarizations.
Therefore, the polarization plane of a plane polarized ray rotates
through the angle
\begin{equation}
\label{11}
\int_C {\bm A}_{\bot}\cdot d{\bm p}=\int_C \cos\theta\,
d\varphi,
\end{equation}
where $\theta (\varphi)$ is the zenith (azimuth) angle in the
spherical polar coordinates of the momentum space. This is the
Rytov-Vladimirskii rotation.

\section{Summery and conclusion}

It was shown that if we consider the post geometric approximation in
unmagnetized plasma media with stationary slowly varying density, an
Abelian gauge field (Berry connection) will result in the ray
Hamiltonian. Appearance of this gauge field leads to an additional
displacement of the photon of distinct helicity in opposite
directions normal to the ray. Also, for a linear polarization, the
rotation of the polarization plane occurs in inhomogeneous
unmagnetized plasmas. This rotation differs from the faraday
rotation which is due to the presence of magnetic field in plasma
media. In fact, it is shown theoretically that the rotation of
polarization plane occurs even in unmagnetized plasma media as a
result of Berry topological phase.

\section{Acknowledgement}
One of the authors (R.T.) wishes to thank to Dr. A. Chakhmachi for
helpful discussions.

%%%%%%%%%%%%%%%%%%%%%%%%%%%%%%%%%%%%%%%%%%%%%%%%%%%%%
\begin{center}
\textbf{Appendix A}
\end{center}
\begin{center}
\textbf{Components of gauge connection}
\end{center}
Direct calculation via (3) yields the following expression for gauge
connection components:
\[
iA_{1} =\left( \begin{array}{ccc} 0 & -\frac{p_y p_z }{p(p_x^2
+p_y^2 )} & \frac{p_y
}{p\sqrt {p_x^2 +p_y^2 }} \\
\frac{p_y p_z }{p(p_x^2 +p_y^2 )} & 0 & -\frac{p_x
 p_z
}{p^2\sqrt {p_x^2 +p_y^2 } } \\
-\frac{p_y }{p\sqrt {p_x^2 +p_y^2 }} & \frac{p_x p_z }{p^2\sqrt
{p_x^2 +p_y^2 }} & 0 \end{array} \right),\;\;
 iA_{2} =\left( {{\begin{array}{ccc}
 0 & {\frac{p_x p_z }{p(p_x^2 +p_y^2 )}} & {-\frac{p_x
}{p\sqrt {p_x^2 +p_y^2 } }} \\
 {-\frac{p_x p_z }{p(p_x^2 +p_y^2 )}} & 0 &
 {-\frac{p_y
 p_z
}{p^2\sqrt {p_x^2 +p_y^2 } }} \\
 {\frac{p_x}{p\sqrt {p_x^2 +p_y^2 } }} & {\frac{p_y p_z }{p^2\sqrt
{p_x^2 +p_y^2 } }} \hfill & 0\\
\end{array} }} \right)
\]
\[
 iA_{3} =\left( {{\begin{array}{ccc}
 0 & 0 & 0 \\
 0 & 0 & {\frac{\sqrt {p_x^2 +p_y^2} }{p^2}} \\
 0 & {-\frac{\sqrt {p_x^2 +p_y^2 }}{p^2}} & 0\\
\end{array}}} \right)
\]


\begin{thebibliography}{widest-label}
\bibitem{Chen} F. F. Chen, \textit{Introduction to plasma physics and controlled fusion}, 2nd ed., Vol. 1 (Plenum Press, New York, 1984).
\bibitem{Hacquin} S. Hacquin, S. Heuraux, M. Colin and G. Leclert, J. Comp. Phys. {\bf 174}, 1 (2001).
\bibitem{Cho} S. Cho, Phys. Plasmas {\bf 11}, 4399 (2004).
\bibitem{Malik} H.K. Malik, S. Kumer and K.P. Singh, Laser Part. Beams {\bf 26}, 197 (2008).
\bibitem{Bliokh}K.Yu. Bliokh and Yu.P. Bliokh, Phys. Rev. E {\bf 70}, 026605 (2004).
\bibitem{Bliokh1} K.Yu. Bliokh and V.D. Freilikher, Phys. Rev. B {\bf 72}, 035108 (2005).
\bibitem{Krall} N.A. Krall and A.W. Trivelpiece, Principles of Plasma Physics (McGraw-Hill, New York, 1973).
\bibitem{Rytov} S.M. Rytov, Dokl. Akad. Nauk SSSR {\bf 18}, 263 (1938).
\bibitem{Vladimirskii} V.V. Vladimirskii, Dokl. Akad. Nauk SSSR {\bf 13}, 222 (1941).
\bibitem{Berry} M.V. Berry, Proc. R. Soc. A {\bf 392}, 45 (1984).
\bibitem{Simon} B. Simon, Phys. Rev. Lett. {\bf 51}, 2167 (1983).
\bibitem{Dooghin} A.V. Dooghin, N.D. Kudnikova, V.S. Liberman and B.Ya. Zel'dovich, Phys. Rev. A {\bf 45}, 8204 (1992).
\bibitem{Liberman} V.S. Liberman, B.Ya. Zel'dovich, Phys. Rev. A {\bf 46}, 5199 (1992).
\bibitem{Bliokh2} K.Yu. Bliokh and Yu.P. Bliokh, Phys. Lett. A {\bf 333}, 181 (2004).
\bibitem{Torabi} R. Torabi and M. Mehrafarin, JETP Lett. {\bf 88}, 590 (2008).
\bibitem{Torabi2} R. Torabi, Can. J. Phys. {\bf 88}, 1 (2010).
\bibitem{Onoda} M. Onoda, S. Murakami and N. Nagaosa, Phys. Rev. Lett. {\bf 93}, 083901 (2004).
\bibitem{Onoda2} M. Onoda, S. Murakami and N. Nagaosa, Phys. Rev. E {\bf 74}, 066610 (2006).
\bibitem{Sawada} K. Sawada, S. Murakami and N. Nagaosa, Phys. Rev. Lett. {\bf 96}, 154802 (2006).
\bibitem{Ginzburg} V. L. Ginzburg, \textit{The Propagation of Electromagnetic Waves in Plasmas} (Gordon and Breach, NY, 1964).
\bibitem{Mehrafarin} M. Mehrafarin and R. Torabi, Phys. Lett. A {\bf 373}, 2114 (2009).
\bibitem{Berard} A. B\'erard and H. Mohrbach, Phys. Lett. A {\bf 352}, 190 (2006).
\bibitem{Kravtsov} YU. A. Kravtsov, \textit{Geometrical Optic of Inhomogeneous Medium} (Springer-Verlag, Berlin, 1990).

\end{thebibliography}
\end{document}